\documentclass[aps, prl, superscriptaddress, reprint]{revtex4-1}
\usepackage{graphicx}  
\usepackage{dcolumn}   
\usepackage{bm}        
\usepackage{amssymb}  
\usepackage{amsfonts}
\usepackage{amsmath}
\usepackage{color}
\usepackage[normalem]{ulem}
\usepackage[percent]{overpic}
\usepackage{soul}
\usepackage{notes2bib}
\usepackage[colorlinks=true,citecolor=blue]{hyperref}
\hypersetup{colorlinks=true,citecolor=blue,linkcolor=red,urlcolor=blue}

\def\be{\begin{equation}}
	\def\ee{\end{equation}}
\def\bs{\begin{split}}
	\def\es{\end{split}}
\def\ber{\begin{eqnarray}}
	\def\eer{\end{eqnarray}}

\def\kv{{\bf k}}
\def\qv{{\bf q}}
\def\pv{{\bf p}}

\begin{document}

\title{Anomalous Lifetime of Quasiparticles in Fermi Liquids as a Precursor of the Density-Wave Instability}
\author{Iran Seydi}
\affiliation{Department of Physics, Institute for Advanced Studies in Basic Sciences (IASBS), Zanjan 45137-66731, Iran }
\author{Saeed H. Abedinpour}
\email{abedinpour@iasbs.ac.ir}
\affiliation{Department of Physics, Institute for Advanced Studies in Basic Sciences (IASBS), Zanjan 45137-66731, Iran }
\author{Reza Asgari}
\affiliation{School of Physics, Institute for Research in Fundamental Sciences (IPM), Tehran 19395-5531, Iran}
\author{B. Tanatar}
\affiliation{Department of Physics, Bilkent University, Bilkent, 06800 Ankara, Turkey}
\date{\today}

\begin{abstract}
We analytically study the inelastic lifetime of quasiparticles due to particle-particle interactions in a three-dimensional Fermi liquid approaching a density-wave instability. Using the G$_0$W approximation, we find that the softening of the dielectric function significantly enhances the quasiparticle decay rate near the instability.
While the zero-temperature quasiparticle lifetime at the Fermi surface generally follows a $|\varepsilon_k-\varepsilon_{\rm F}|^{-\alpha}$ divergence with $\alpha=2$, we observe $\alpha=0.5$ at the instability point and $\alpha=1$ within the density-wave phase.
Moreover, we demonstrate that the renormalization constant $Z$ is substantially suppressed as the instability is approached, enhancing the effective mass. 
We extend our analysis to ultra-cold Rydberg-dressed Fermi liquids, where the soft-core interactions promote density-wave instability, and find that our numerical G$_0$W results are in excellent agreement with our analytic predictions for quasiparticle lifetime, renormalization constant, and effective mass.
\end{abstract}
\maketitle

\emph{Introduction.}
Landau's theory of Fermi liquids imposes general constraints on the behavior of quasiparticles in interacting homogeneous Fermi systems. According to this theory, at zero temperature, the inverse lifetime of a quasiparticle with wave-vector $k$ should vanish as $(k-k_{\rm F})^2$ in three dimensions~\cite{Wolfle_RPP2018,Vignale2005,Qian2005}.
While the Landau Fermi liquid (FL) model successfully explains most experimental results where low-energy excitations are weakly interacting quasiparticles, various alternative phases have been theoretically predicted and experimentally sought. These alternative phases can be broadly classified as homogeneous or non-homogeneous, depending on whether translational invariance is preserved or broken, as seen in Bloch ferromagnetic and BCS superconducting phases.
Much less is known about the behavior of quantum many-body systems at the onset of, and within, a density-wave instability (DWI), where periodic modulations in particle density emerge due to inter-particle interactions~\cite{Vignale2005,baym2008}. However, having long-lived quasiparticles is generally a crucial prerequisite for defining quasiparticles in the first place~\cite{Hartnoll_Book2018}. 

An intriguing aspect of quasiparticle lifetime at the onset of DWI is its nontrivial dependence on momentum. Near the Fermi surface, where states are most sensitive to perturbations, quasiparticle lifetime shows significant deviations from conventional FL behavior, often manifested as anomalous scattering rates linked to the density-wave order. Additionally, the influence of quantum criticality, as the system approaches a continuous phase transition at zero temperature, is significant. Quasiparticles may become increasingly long-range correlated near the critical point, leading to enhanced quantum fluctuations.

Understanding quasiparticle behavior as a FL approaches instability is key to unraveling the underlying physics~\cite{Debbeler2024,Miserev2023,Metlitski2010,Altshuler1995,Bergeron2012}. Central to this investigation is the characteristic timescale of the quasiparticle lifetime, which plays a crucial role in the transport phenomena of these systems~\cite{Berk1995}.

In recent years, experimental and theoretical efforts have aimed to elucidate the behavior of quasiparticles near the critical point of the density-wave transition. Techniques such as angle-resolved photoemission spectroscopy and scanning tunneling microscopy have provided invaluable data on the evolution of electronic states and quasiparticle lifetimes in various materials~\cite{Murphy1995,Berk1995,Slutzky1996,bostwick2010}. Theoretical frameworks, including renormalization group theory and quantum many-body numerical simulations, have offered insights into the microscopic processes that govern quasiparticle dynamics in different FL systems~\cite{Lee2018,Metlitski2010,Menashe1996,Qian2005, Seydi_PRA2018,Debbeler2024,Gerlach2017}. 
Recent experimental progress in cooling and trapping ultra-cold atoms and molecules with long-range interactions, such as dipolar and Rydberg-dressed gases, has also provided a new platform for the inspection of the properties of the density-wave phase~\cite{Recati_NRP2023, Han_PRL2018, Zhu_PRR2023}.

The dynamics of quasiparticles at the onset of DWI provide valuable insights into the mechanisms driving this phase transition~\cite{Song2023}. Non-Fermi liquid behavior has been studied in two-dimensional (2D) metals near the onset of incommensurate $2k_{\rm F}$ charge or spin-density wave order. At ``hot spots", the single-particle decay rate, calculated using perturbative one-loop methods, follows a power-law dependence on energy with exponents of $2/3$ and $1$ for single and two hot spot pairs, respectively~\cite{Holder2014}. Instabilities of the FL near magnetic quantum phase transitions have also been explored~\cite{Lohneysen2007}.
Marginal FL behavior has been observed at the onset of $2k_{\rm F}$ density-wave order in 2D metals with flat hot spots, and logarithmic vanishing of quasiparticle weight and Fermi velocity predicted at the hot spots~\cite{Debbeler2024}.

The energy dependence of quasiparticle decay rates in geometric quantum critical metals, where the Fermi surface has inflection points, has been studied too~\cite{Song2023}. Near these inflection points, where the local curvature of the Fermi surface vanishes, the quasiparticle lifetime near the Fermi surface in a two-dimensional geometric quantum critical metal follows $\varepsilon^{\alpha/(\alpha-1)}$ for $\alpha > 2$, where $\alpha$ is the order of the inflection points~\cite{Song2023}.

In this work, we analytically examine the behavior of quasiparticle lifetime in a three-dimensional (3D) FL at zero temperature as the system approaches the DWI. 
The DWI we have in mind is a purely many-body effect and is distinct from the charge density phase in materials usually originating in or accompanied by a periodic distortion of the lattice~\cite{Gruner_RMP1988,Zhu_JPCS2022}.

Using the G$_0$W approximation, we find that the decay rate is significantly enhanced near the instability point, even while the system remains in the FL regime. This enhancement is attributed to the softening of the excitation spectrum, leading to the emergence of a roton-like feature~\cite{Bohm_PRB2010,Godfrin2012}. At the threshold between the FL and density-wave phases, the decay rate vanishes with an unusually slow slope, exhibiting a momentum dependence of $|k-k_{\rm F}|^{0.5}$, where $k_{\rm F}$ is the Fermi wave vector.
Extending our calculations based on the self-energy of a homogeneous Fermi system into the density-wave (DW) phase -- despite the inherent limitations of this approach beyond the normal Fermi liquid (NFL) phase-- we predict a decay rate that vanishes as $|k-k_{\rm F}|$. This decay is slower than in the FL regime but faster than at the instability point.

We also examine the behavior of the quasiparticle renormalization factor and the NFL's effective mass at the onset of DWI. We find strong suppression of the renormalization factor and enhancement of the effective mass as the system approaches the instability point.

To verify the accuracy of these results, we investigate the Landau-Fermi liquid properties of ultra-cold Rydberg-dressed FL using the G$_0$W approximation. The soft-core nature of the Rydberg-dressed interactions drives the system toward the DWI, and even a simple random-phase approximation (RPA) can capture this instability~\cite{khasseh2017phase, Seydi2021}. Our numerical results for the quasiparticle properties in various regimes align perfectly with our analytic findings.

\emph{Model.}
Within the G$_0$W approximation, we write the quasiparticle self-energy as~\cite{Vignale2005}
\begin{equation}\label{G0_W}
	\Sigma(k,E)= i  
	 \int\frac{\mathrm{d}^{\rm 3}{\bf q}\,\mathrm{d}(\hbar\omega)}{(2\pi)^{4}}
	{\cal G}^{0}(\textbf{k}-\textbf{q},E-\hbar \omega) W(q,\omega),
\end{equation}  
where 
${\cal G}^{0}(\textbf{k}, E) = 1/(E+i0^+ - \varepsilon_k)$ is the non-interacting retarded Green's function with $\varepsilon_k=\hbar^2k^2/(2m)$ the bare particle dispersion, and $W(q,\omega)$ is the effective particle-particle interaction. In the Kukkonen-Overhauser approximation, we can express the effective interaction as~\cite{Vignale2005}
$W_{\rm KO}(q,\omega) = v(q) + w^2(q,\omega) \chi(q,\omega)$,
where $v(q)$ is the Fourier transform of the bare particle-particle interaction, and $\chi (q,\omega)$ is the interacting density-density linear response function which we can express in terms of the non-interacting density-density linear response (Lindhard function) $\chi_{0} (q,\omega)$, and dynamically screened interaction $w(q,\omega)$, as
\begin{equation}\label{response}
	\chi (q,\omega)=\frac{\chi_{0} (q,\omega)}{1-w(q,\omega)\chi_{0} (q,\omega)}.
\end{equation} 
Note that the effects of the exchange-correlation hole are included in the screened interaction, and it is usually expressed in terms of the dynamical local field factor (LFF) $G(q,\omega)$, as $w(q,\omega)=v(q)[1-G(q,\omega)]$~\cite{Vignale2005}. 
In the celebrated RPA, one replaces the screened interaction with bare one $v(q)$ (i.e., entirely discards the LFF). Even in more elaborate approximations, the frequency dependence of the LFF factor is often dropped, and the screened interaction is approximated with a static and real potential $w(q)$. 

\emph{Quasiparticle lifetime.}
It is customary to split the G$_0$W self-energy~Eq.~\eqref{G0_W} into Fock, line, and pole terms~\cite{Seydi_PRA2018}, where only the last one contributes to the imaginary part of the self-energy, that in the on-shell approximation, i.e., $E\to \xi_k$, and after making the change of variables $\pv=\kv-\qv$, reads
\be\label{sigma_im_OSA}
\begin{split}
	{\rm Im}\,\Sigma (k,\xi_k) &=  
	\int\frac{\mathrm{d}^{3}{\pv}}{(2\pi)^3}
	w^{2}(|\kv-\pv|){\rm Im}\,\chi(|\kv-\pv|,\xi_k-\xi_{p}) \\
& \quad	\times \left[\Theta(\xi_k-\xi_p)-\Theta(-\xi_p)\right].
\end{split}
\ee
Here, $\xi_k=\varepsilon_k-\varepsilon_{\rm F}$, where $\varepsilon_{\rm F}=\hbar^2 k^2_{\rm F}/(2 m)$ is the Fermi energy. 
The imaginary part of the density-density response function, when a static screened interaction is adopted, is non-zero only inside the particle-hole continuum (PHC), as well as right on the collective-mode dispersion, i.e., $\omega= \Omega_{\rm col.}(q)$~\cite{Vignale2005}.

\begin{figure}
	\centering 	
	\includegraphics[width=\linewidth]{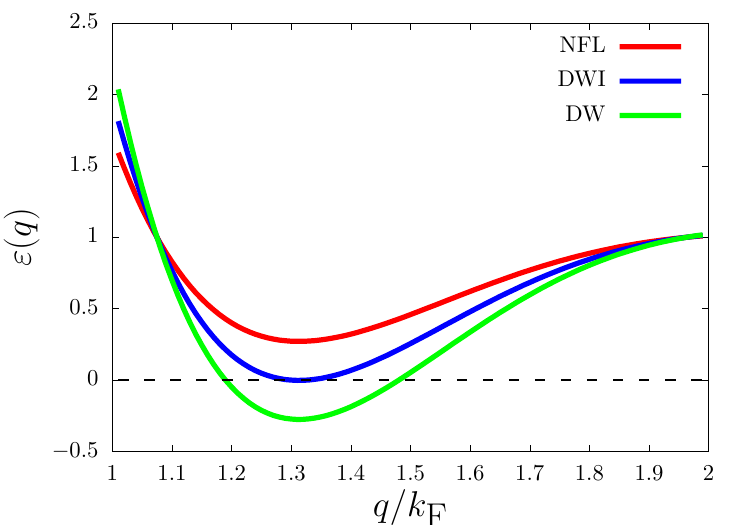}
	\caption{An schematic scratch of the static dielectric function versus $q/k_{\rm F}$ in different regimes. In the normal Fermi liquid (NFL) regime, the dielectric function remains positive for all ranges of the wave vectors. 
	When $\epsilon(q)$ vanishes for $q=q_c$, the homogeneous becomes unstable to density waves (DWI).
	In the density wave (DW) regime, the dielectric function of the homogeneous Fermi gas becomes negative for $q_{c,1}<q<q_{c,2}$. 
		\label{fig:dielectric_schematic}} 
\end{figure}

In the following, we will investigate the behavior of quasiparticle lifetime $\tau_k=-\hbar/{\rm Im}\,\Sigma (k,\xi_k)$  close to the Fermi level, in a pristine 3D FL at vanishing temperatures. 
We consider three different regimes:
(i) Homogeneous normal FL,
(ii) at the DWI point (i.e., the boundary between FL and density-wave phase), and (iii)
deep inside the DW phase. 
We identify these three regimes from the behavior of static dielectric function $\epsilon(q)=1-w(q)\chi_{0} (q)$. 
Large particle density fluctuations occur at low frequency when $\epsilon(q,\omega=0)$ tends to zero. 
For a normal FL, the static dielectric function remains positive for the entire range of wave vectors.
The dielectric function becomes soft with a positive minimum at a finite $q$ as the system moves towards the DWI. At the instability point, $\epsilon(q)$ vanishes at wave vector $q_c$, corresponding to the instability wavelength $\lambda_c=2\pi/q_c$. Inside the DW phase, $\epsilon(q)$ becomes negative for a range of wave vectors, i.e., $q_{c,1}<q<q_{c,2}$ (see, Fig.~\ref{fig:dielectric_schematic} for an illustration of dielectric function in different regimes). 
Notice that $1-w(q)\chi_{0} (q)$ is the dielectric function of the \emph{normal FL}, and not the one of the DW phase, as the actual static dielectric function should remain positive.

It is straightforward to demonstrate that collective mode outside the PHC does not contribute to the inelastic scattering of quasiparticles in the vicinity of the Fermi surface, and all the decay is due to inelastic scatterings within the PHC \cite{Note2}. 
For normal Fermi liquids, the static dielectric function is positive, and we can easily evaluate the $k\to k_{\rm F}$ limit in Eq. \eqref{sigma_im_OSA} to find the well-established behavior of the quasiparticle lifetime~\cite{Vignale2005, Qian2005}
\be\label{eq:tau_FL}
\frac{\hbar}{\tau^{\rm FL}_{k\to k_{\rm F}}}
=
\alpha \,\varepsilon_{\rm F} \left|\frac{k}{k_{\rm F}}-1\right|^2.
\ee
Here, dimensionless coefficient $\alpha$ is defined as
\be\label{eq:alpha3}
\alpha=\frac{\pi}{4} \int_{0}^{2} \mathrm{d}y \frac{\nu_0^2 w^{2}(y)}{\left|1-w(y) \chi_0(y)\right|^2},
\ee
where $y=q/k_{\rm F}$, and $\nu_0=m k_{\rm F}/(2\pi^2 \hbar^2)$ is the density of states at the Fermi level of a spinless 3D ideal Fermi gas~\cite{Vignale2005}.

As the system moves toward the DWI, the static dielectric function becomes softer and approaches zero at a finite wave vector. Close to the instability region, the $1-w(q) \chi_0(q)$ factor in the denominator of Eq.~\eqref{eq:alpha3} enhances $\alpha$. This suppression of the quasiparticle lifetime stems from the extra weight that scattering into the soft roton mode of strongly interacting FL carries~\cite{Godfrin2012, Seydi2021}.

Right at the DWI point, where $1-w(q) \chi_0(q)$ becomes zero at a finite wave vector (see the blue curve in Fig.~\ref{fig:dielectric_schematic}), handling the singular behavior of integrand in Eq.~\eqref{sigma_im_OSA} requires extra care. Expanding $1-w(q)\chi_0(q,\omega)$ around $(q\to q_c,\omega\to0)$, we find contributions to the imaginary part of the self-energy that vanish much slower than the FL regime, i.e.,
\be\label{eq:tau_inst}
\frac{\hbar}{\tau^{\rm inst.}_{k\to k_{\rm F}}}
=
\beta \,\varepsilon_{\rm F} \sqrt{\left|\frac{k}{k_{\rm F}}-1\right|}.
\ee
with the dimensionless coefficient of the instability point $\beta$ defined as
\be\label{eq:beta3}
\beta=\sqrt{\frac{8 \pi y_c^3 \nu_0 w(y_c)}{\partial^2_y\left.[w(y) \chi_{0} (y)]\right|_{y=y_c}}},
\ee 
where $\partial_y\equiv \partial/\partial_y$, and $y_c=q_c/k_{\rm F}$ is the dimensionless instability wave vector.

As we pass to the DW phase, the static dielectric function obtained from the polarizability of the homogeneous FL, 
becomes negative for $q_{c,1}<q<q_{c,2}$ (see, the green curve in Fig.~\ref{fig:dielectric_schematic}).
The homogeneous FL is no longer the system's true ground state, but we keep working with it to see what it predicts for the quasiparticle lifetime in the DW regime.
Now, expanding $1-w(q)\chi_0(q,\omega)$ around $(q\to q_{c,j},\omega\to0)$, where $j=1,2$, again we find that the imaginary part of the self-energy vanishes much slower than the FL regime, but faster than the instability point \cite{Note2}
\be\label{eq:tau_DW}
\frac{\hbar}{\tau^{\rm DW}_{k\to k_{\rm F}}}
=
\gamma \,\varepsilon_{\rm F} {\left|\frac{k}{k_{\rm F}}-1\right|},
\ee
where the dimensionless $\gamma$ coefficient for the DW regime is
\be\label{eq:gamma3}
\gamma= \pi  \sum_{j=1,2} \left|\frac{y_{c,j}\nu_0w(y_{c,j})}{\left.\partial_y[w(y) \chi_{0} (y)]\right|_{y=y_{c,j}}} \right|.
\ee
Here, $y_{c,j}=q_{c,j}/k_{\rm F}$, refer to two (dimensionless) wave vectors where the static dielectric function vanishes  (see, the green curve in Fig.~\ref{fig:dielectric_schematic}).

\emph{Renormalization constant and effective mass.}
The renormalization constant
$Z^{-1}=1-\left.\partial_E {\rm Re}\,\Sigma (k_{\rm F},E) \right|_{E=0}$,
is the discontinuity of the zero-temperature momentum distribution at the Fermi level is generally reduced by interactions. However, the Landau Fermi liquid picture implies a non-vanishing $Z$~\cite{Vignale2005}. 
The pole contribution to the renormalization factor in the FL regime is
\be\label{eq:Z_pole}
\left.\partial_E {\rm Re}\,\Sigma_{\rm pole} (k_{\rm F},E) \right|_{E=0}=
\frac{\nu_0}{2} \int_{0}^{2} \mathrm{d}y y\frac{ w^{2}(y) \chi_0(y)}{1-w(y) \chi_0(y)}.
\ee
As we approach the instability point from the FL phase, the static dielectric function in the denominator becomes softer and $\partial_E {\rm Re}\,\Sigma_{\rm pole} (k_{\rm F},E)$ becomes very large, 
and it may even diverge right at the instability point where the dielectric function vanishes.
The renormalization constant, therefore, is suppressed as  
\be\label{eq:Z_DWI}
Z \approx \frac{-1}{\left.\partial_E {\rm Re}\,\Sigma_{\rm pole} (k_{\rm F},E) \right|_{E=0}},
\ee
as we approach the DWI.

The suppression of the renormalization constant gives rise to the enhancement of the effective mass
\be\label{eq:m*}
\frac{m}{m^*}=Z\left[1+\frac{m}{\hbar^2 k_{\rm F}}\left.\partial_k {\rm Re}\,\Sigma (k,0) \right|_{k=k_{\rm F}}\right],
\ee
as the system approaches DWI. 
We should note that such instability has also been studied in electronic systems~\cite{Khodel_JETPLett1997, Yakovenko_JETPLett2003,Zhang_PRB2005,Asgari_PRB2009}, and is a generic feature of strongly correlated electron liquids with $1/r$ Coulomb interaction.

Eqs.~\eqref{eq:tau_FL}-\eqref{eq:gamma3} and Eq.~\eqref{eq:Z_DWI} constitute the main results of our work. In what follows, we put our findings on the quasiparticle lifetime, renormalization constant, and effective mass to test in a Fermi gas with Rydberg-dressed interactions that are unstable to density-wave instability at strong interactions. 

\emph{Soft-core interaction and numerical results.}
Now we consider a Fermi gas with Rydberg-dressed interaction 
\be\label{eq:RD}
v_{\rm RD}(r)=\frac{U}{1+(r/R_c)^6},
\ee 
where $U$ and $R_c$ are the interaction strength and soft-core radius of the interaction, respectively. 
We can specify the zero temperature behavior of this system with two dimensionless parameters $u=U/\varepsilon_{\rm F}$, and $r_c=R_c k_{\rm F}$~\cite{Seydi2021}.
Due to the soft-core form of the Rydberg-dressed interaction, its Fourier transform has a negative minimum at a finite wave vector $q$. 
For the sake of simplicity, we employ the RPA to calculate the QP lifetime.
Already at the RPA level, the static dielectric function of Rydberg-dressed gas can become negative at strong interactions and large soft-core radii. 
Such behavior indicates the instability of the homogeneous Fermi gas to density-modulated phases~\cite{Seydi2021,khasseh2017phase}. 
Within the RPA, the DW instability is expected for $u \geq 5.49$ and $u \geq2.62$ at $r_c=4$ and $r_c=5$,  respectively~\cite{Seydi2021}.

\begin{figure}
	\centering 	
		\includegraphics[width=\linewidth]{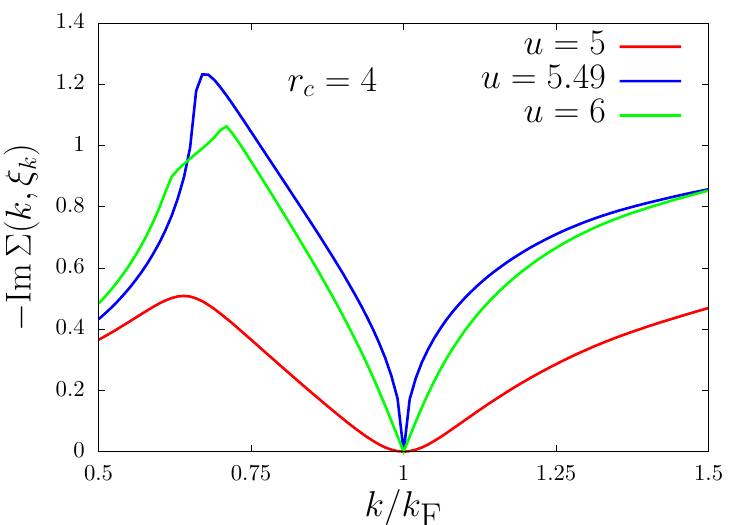}
	\caption{The behavior of ${\rm Im}\,\Sigma(k\to k_{\rm F}, \xi_k\to 0)$ (in the units of $\varepsilon_{\rm F}$) of a  3D  Rydberg dressed gas for $r_c=4$ and varying interaction strengths $u$, corresponding to the homogeneous FL, instability region, and density-wave phase. For $r_c=4$, the instability manifests itself for $u \geq 5.49$.
		\label{fig:sigma}} 
\end{figure}

We illustrate the imaginary part of the self-energy near the Fermi level across different regimes in Fig.~\ref{fig:sigma}. The different scaling behavior of self-energy is evident from the plots.

\begin{figure*}
	\centering 	
	\includegraphics[width=\linewidth]{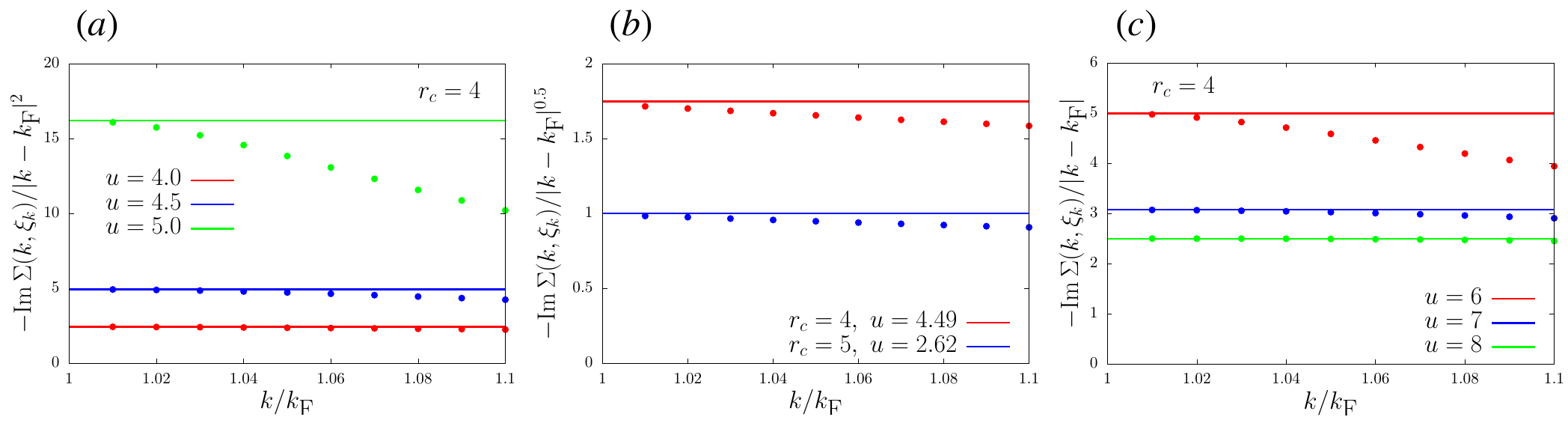}
	\caption{The behavior of ${\rm Im}\,\Sigma(k\to k_{\rm F}, \xi_k\to 0)$ of a three dimensional Rydberg dressed Fermi gas in the homogeneous FL regime $(a)$,
	right at the density-wave instability point $(b)$ and inside the density-wave phase $(c)$ for different interaction strength and soft-core radius values.
		Filled symbols are the numerical data obtained directly from Eq.~\eqref{sigma_im_OSA} for the imaginary part of the self-energy. The solid lines show the values for the coefficients $\alpha$, $\beta$, and $\gamma$ from Eqs.~\eqref{eq:alpha3},~\eqref{eq:beta3}, and~\eqref{eq:gamma3}, respectively. 
		\label{fig:sigma_scaling}} 
\end{figure*}

For a normal FL, we have ${\rm Im}\,\Sigma(k\to k_{\rm F},\xi_k\to 0) \propto   |k-k_{\rm F}|^2$.  In the right panel of Fig.~\ref{fig:sigma_scaling}, we compare the behavior of ${\rm Im}\, \Sigma$, obtained from the solution of G$_0$W equation [i.e., Eq.~\eqref{sigma_im_OSA}] with the semi-analytic results we find for the coefficient $\alpha$ from Eq. \eqref{eq:alpha3}.
At the density-wave instability point, we expect the imaginary part of the self-energy at the Fermi surface to behave as $ \propto \sqrt{|k-k_{\rm F}|}$.
The middle panel in Fig.~\ref{fig:sigma_scaling} illustrates this behavior for two points at the phase transition line between the FL and density-wave phases.
Finally, once the system is in the density-wave phase, the imaginary part of the self-energy (calculated from the RPA for the homogeneous system) vanishes as $\propto |k-k_{\rm F}|$. We show this behavior in the right panel of Fig.~\ref{fig:sigma_scaling}.

Fig.~\ref{fig:sigma_scaling} confirms that our analytically anticipated behavior of the quasiparticle lifetime coincides with the numerical results for the 3D FL with Rydberg-dressed interaction across different regimes.

\begin{figure}
	\centering 	
		\includegraphics[width=\linewidth]{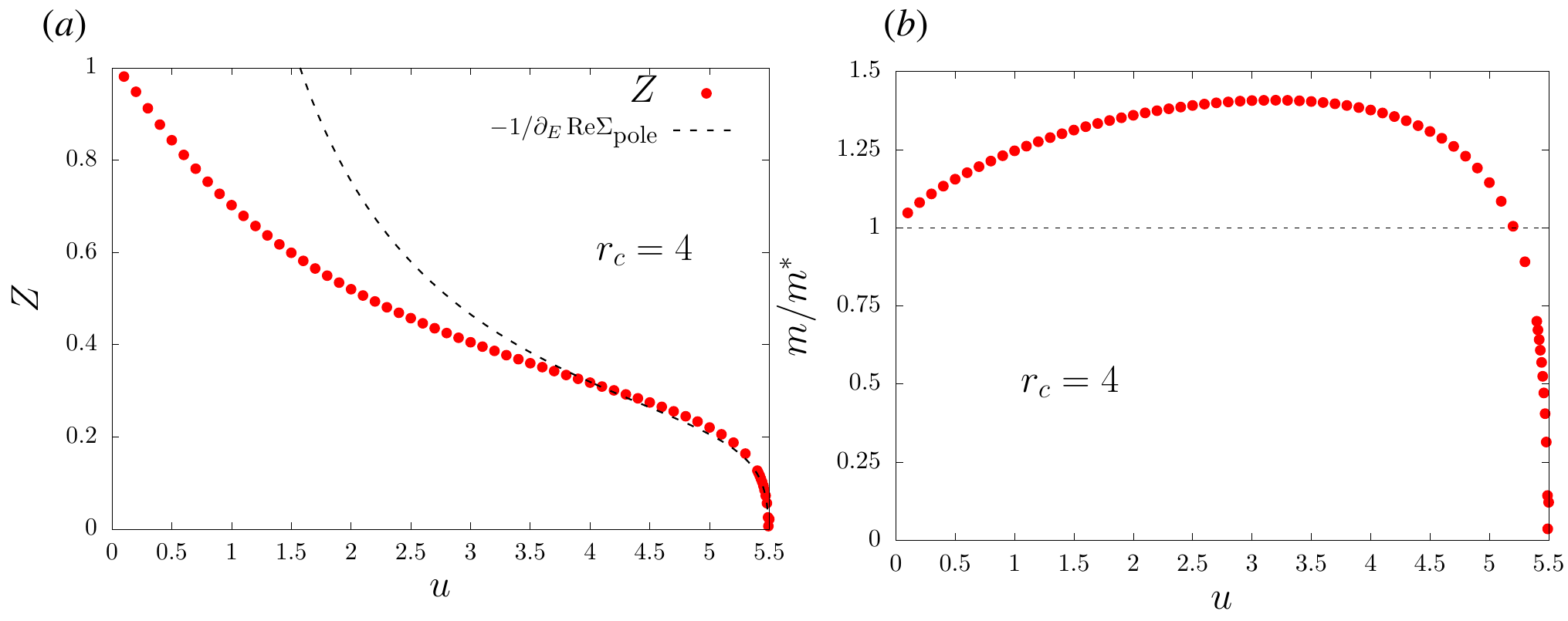}
	\caption{The behavior of renormalization factor $Z$ $(a)$ and inverse effective mass $m^*$ $(b)$, in the units of bare mass $m$, of a  3D  Rydberg dressed gas  versus interaction strength $u$ for $r_c=4$. 
	The dashed line in panel $(a)$ shows the analytic results for the contribution of the pole term to the renormalization factor obtained from Eq.~\eqref{eq:Z_pole}.
	The instability emerges at $u \approx 5.49$ for  $r_c=4$.
		\label{fig:Z_mass}} 
\end{figure}

In Fig.~\ref{fig:Z_mass}, we show the behavior of renormalization constant and effective mass of a Rydberg-dressed FL in its normal regime. For $r_c=4$, the renormalization factor vanishes at $u\approx 5.49$ where the DWI emerges, and its behavior close to the instability is well captured by our semi-analytic expression~\eqref{eq:Z_DWI}.
As expected, the numerical results show the suppression of effective mass at small interactions, where the Hartree-Fock effects are dominant~\cite{Vignale2005}. At strong interactions, the dynamic contributions to self-energy dominate, and eventually, the effective mass diverges at the onset of the DWI.

\emph{Conclusions.}
In summary, we used the G$_0$W approximation to study the quasiparticle properties of a FL approaching the DWI.
Near the Fermi surface, quasiparticles scatter only within the particle-hole continuum, where energy and momentum conservation can be satisfied. At strong interactions, a roton-like minimum appears in the dispersion of collective modes within the particle-hole continuum, indicating a tendency towards density-modulated phases. This softening of the collective mode enhances the quasiparticle decay rate even within the FL phase, though the lifetime still vanishes according to the Landau FL paradigm~\cite{Vignale2005}.

As the Fermi system becomes unstable to density-modulated phases, the quasiparticle lifetime diverges at the Fermi level but with revised scaling behavior. The slowest divergence occurs at the boundary between homogeneous and DW phases.
Other quasiparticle properties also signal the DWI in a strongly correlated FL. This includes a vanishing renormalization constant and a diverging effective mass at the instability point.

We also analyzed the imaginary part of the self-energy and other quasiparticle properties in Rydberg-dressed gases using the Fermi hyper-netted chain approach~\cite{Seydi2021,Iran_quasiparticle_draft}, observing similar scaling behavior for different values of $r_c$ and $u$.

\emph{Acknowledgements.}  
We are grateful to Giovanni Vignale for his valuable insights. 
The work in Zanjan is supported by the Research Council of the Institute for Advanced Studies in Basic Sciences (IASBS).
BT is supported by the Scientific and Technological Research Councl of T{\"u}rkiye (TUBITAK) under Grant No.\,119N689
and the Turkish Academy of Sciences (TUBA).

\bibliography{QP_DWI.bbl}

\begin{thebibliography}{36}%
\makeatletter
\providecommand \@ifxundefined [1]{%
 \@ifx{#1\undefined}
}%
\providecommand \@ifnum [1]{%
 \ifnum #1\expandafter \@firstoftwo
 \else \expandafter \@secondoftwo
 \fi
}%
\providecommand \@ifx [1]{%
 \ifx #1\expandafter \@firstoftwo
 \else \expandafter \@secondoftwo
 \fi
}%
\providecommand \natexlab [1]{#1}%
\providecommand \enquote  [1]{``#1''}%
\providecommand \bibnamefont  [1]{#1}%
\providecommand \bibfnamefont [1]{#1}%
\providecommand \citenamefont [1]{#1}%
\providecommand \href@noop [0]{\@secondoftwo}%
\providecommand \href [0]{\begingroup \@sanitize@url \@href}%
\providecommand \@href[1]{\@@startlink{#1}\@@href}%
\providecommand \@@href[1]{\endgroup#1\@@endlink}%
\providecommand \@sanitize@url [0]{\catcode `\\12\catcode `\$12\catcode
  `\&12\catcode `\#12\catcode `\^12\catcode `\_12\catcode `\%12\relax}%
\providecommand \@@startlink[1]{}%
\providecommand \@@endlink[0]{}%
\providecommand \url  [0]{\begingroup\@sanitize@url \@url }%
\providecommand \@url [1]{\endgroup\@href {#1}{\urlprefix }}%
\providecommand \urlprefix  [0]{URL }%
\providecommand \Eprint [0]{\href }%
\providecommand \doibase [0]{http://dx.doi.org/}%
\providecommand \selectlanguage [0]{\@gobble}%
\providecommand \bibinfo  [0]{\@secondoftwo}%
\providecommand \bibfield  [0]{\@secondoftwo}%
\providecommand \translation [1]{[#1]}%
\providecommand \BibitemOpen [0]{}%
\providecommand \bibitemStop [0]{}%
\providecommand \bibitemNoStop [0]{.\EOS\space}%
\providecommand \EOS [0]{\spacefactor3000\relax}%
\providecommand \BibitemShut  [1]{\csname bibitem#1\endcsname}%
\let\auto@bib@innerbib\@empty
\bibitem [{\citenamefont {W\"{o}lfle}(2018)}]{Wolfle_RPP2018}%
  \BibitemOpen
  \bibfield  {author} {\bibinfo {author} {\bibfnamefont {P.}~\bibnamefont
  {W\"{o}lfle}},\ }\href {\doibase 10.1088/1361-6633/aa9bc4} {\bibfield
  {journal} {\bibinfo  {journal} {Reports on Progress in Physics}\ }\textbf
  {\bibinfo {volume} {81}},\ \bibinfo {pages} {032501} (\bibinfo {year}
  {2018})}\BibitemShut {NoStop}%
\bibitem [{\citenamefont {Giuliani}\ and\ \citenamefont
  {Vignale}(2005)}]{Vignale2005}%
  \BibitemOpen
  \bibfield  {author} {\bibinfo {author} {\bibfnamefont {G.~F.}\ \bibnamefont
  {Giuliani}}\ and\ \bibinfo {author} {\bibfnamefont {G.}~\bibnamefont
  {Vignale}},\ }\href@noop {} {\emph {\bibinfo {title} {Quantum Theory of the
  Electron Liquid}}},\ Masters Series in Physics and Astronomy\ (\bibinfo
  {publisher} {Cambridge University Press},\ \bibinfo {year}
  {2005})\BibitemShut {NoStop}%
\bibitem [{\citenamefont {Qian}\ and\ \citenamefont
  {Vignale}(2005)}]{Qian2005}%
  \BibitemOpen
  \bibfield  {author} {\bibinfo {author} {\bibfnamefont {Z.}~\bibnamefont
  {Qian}}\ and\ \bibinfo {author} {\bibfnamefont {G.}~\bibnamefont {Vignale}},\
  }\href {\doibase 10.1103/PhysRevB.71.075112} {\bibfield  {journal} {\bibinfo
  {journal} {Phys. Rev. B}\ }\textbf {\bibinfo {volume} {71}},\ \bibinfo
  {pages} {075112} (\bibinfo {year} {2005})}\BibitemShut {NoStop}%
\bibitem [{\citenamefont {Baym}\ and\ \citenamefont
  {Pethick}(2008)}]{baym2008}%
  \BibitemOpen
  \bibfield  {author} {\bibinfo {author} {\bibfnamefont {G.}~\bibnamefont
  {Baym}}\ and\ \bibinfo {author} {\bibfnamefont {C.}~\bibnamefont {Pethick}},\
  }\href@noop {} {\emph {\bibinfo {title} {Landau Fermi-Liquid Theory: Concepts
  and Applications}}}\ (\bibinfo  {publisher} {Wiley},\ \bibinfo {year}
  {2008})\BibitemShut {NoStop}%
\bibitem [{\citenamefont {Hartnoll}\ \emph {et~al.}(2018)\citenamefont
  {Hartnoll}, \citenamefont {Lucas},\ and\ \citenamefont
  {Sachdev}}]{Hartnoll_Book2018}%
  \BibitemOpen
  \bibfield  {author} {\bibinfo {author} {\bibfnamefont {S.~A.}\ \bibnamefont
  {Hartnoll}}, \bibinfo {author} {\bibfnamefont {A.}~\bibnamefont {Lucas}}, \
  and\ \bibinfo {author} {\bibfnamefont {S.}~\bibnamefont {Sachdev}},\
  }\href@noop {} {\emph {\bibinfo {title} {Holographic Quantum Matter}}},\ The
  MIT Press\ (\bibinfo  {publisher} {MIT Press},\ \bibinfo {year}
  {2018})\BibitemShut {NoStop}%
\bibitem [{\citenamefont {Debbeler}\ and\ \citenamefont
  {Metzner}(2024)}]{Debbeler2024}%
  \BibitemOpen
  \bibfield  {author} {\bibinfo {author} {\bibfnamefont {L.}~\bibnamefont
  {Debbeler}}\ and\ \bibinfo {author} {\bibfnamefont {W.}~\bibnamefont
  {Metzner}},\ }\href {\doibase 10.1103/PhysRevB.109.235112} {\bibfield
  {journal} {\bibinfo  {journal} {Phys. Rev. B}\ }\textbf {\bibinfo {volume}
  {109}},\ \bibinfo {pages} {235112} (\bibinfo {year} {2024})}\BibitemShut
  {NoStop}%
\bibitem [{\citenamefont {Miserev}\ \emph {et~al.}(2023)\citenamefont
  {Miserev}, \citenamefont {Schoeller}, \citenamefont {Klinovaja},\ and\
  \citenamefont {Loss}}]{Miserev2023}%
  \BibitemOpen
  \bibfield  {author} {\bibinfo {author} {\bibfnamefont {D.}~\bibnamefont
  {Miserev}}, \bibinfo {author} {\bibfnamefont {H.}~\bibnamefont {Schoeller}},
  \bibinfo {author} {\bibfnamefont {J.}~\bibnamefont {Klinovaja}}, \ and\
  \bibinfo {author} {\bibfnamefont {D.}~\bibnamefont {Loss}},\ }\href@noop {}
  {\enquote {\bibinfo {title} {Microscopic mechanism of pair-, charge- and
  spin-density-wave instabilities in interacting d-dimensional fermi
  liquids},}\ } (\bibinfo {year} {2023}),\ \Eprint
  {http://arxiv.org/abs/2312.17208} {arXiv:2312.17208 [cond-mat.str-el]}
  \BibitemShut {NoStop}%
\bibitem [{\citenamefont {Metlitski}\ and\ \citenamefont
  {Sachdev}(2010)}]{Metlitski2010}%
  \BibitemOpen
  \bibfield  {author} {\bibinfo {author} {\bibfnamefont {M.~A.}\ \bibnamefont
  {Metlitski}}\ and\ \bibinfo {author} {\bibfnamefont {S.}~\bibnamefont
  {Sachdev}},\ }\href {\doibase 10.1103/PhysRevB.82.075128} {\bibfield
  {journal} {\bibinfo  {journal} {Phys. Rev. B}\ }\textbf {\bibinfo {volume}
  {82}},\ \bibinfo {pages} {075128} (\bibinfo {year} {2010})}\BibitemShut
  {NoStop}%
\bibitem [{\citenamefont {Altshuler}\ \emph {et~al.}(1995)\citenamefont
  {Altshuler}, \citenamefont {Ioffe},\ and\ \citenamefont
  {Millis}}]{Altshuler1995}%
  \BibitemOpen
  \bibfield  {author} {\bibinfo {author} {\bibfnamefont {B.~L.}\ \bibnamefont
  {Altshuler}}, \bibinfo {author} {\bibfnamefont {L.~B.}\ \bibnamefont
  {Ioffe}}, \ and\ \bibinfo {author} {\bibfnamefont {A.~J.}\ \bibnamefont
  {Millis}},\ }\href {\doibase 10.1103/PhysRevB.52.5563} {\bibfield  {journal}
  {\bibinfo  {journal} {Phys. Rev. B}\ }\textbf {\bibinfo {volume} {52}},\
  \bibinfo {pages} {5563} (\bibinfo {year} {1995})}\BibitemShut {NoStop}%
\bibitem [{\citenamefont {Bergeron}\ \emph {et~al.}(2012)\citenamefont
  {Bergeron}, \citenamefont {Chowdhury}, \citenamefont {Punk}, \citenamefont
  {Sachdev},\ and\ \citenamefont {Tremblay}}]{Bergeron2012}%
  \BibitemOpen
  \bibfield  {author} {\bibinfo {author} {\bibfnamefont {D.}~\bibnamefont
  {Bergeron}}, \bibinfo {author} {\bibfnamefont {D.}~\bibnamefont {Chowdhury}},
  \bibinfo {author} {\bibfnamefont {M.}~\bibnamefont {Punk}}, \bibinfo {author}
  {\bibfnamefont {S.}~\bibnamefont {Sachdev}}, \ and\ \bibinfo {author}
  {\bibfnamefont {A.-M.~S.}\ \bibnamefont {Tremblay}},\ }\href {\doibase
  10.1103/PhysRevB.86.155123} {\bibfield  {journal} {\bibinfo  {journal} {Phys.
  Rev. B}\ }\textbf {\bibinfo {volume} {86}},\ \bibinfo {pages} {155123}
  (\bibinfo {year} {2012})}\BibitemShut {NoStop}%
\bibitem [{\citenamefont {Berk}\ \emph {et~al.}(1995)\citenamefont {Berk},
  \citenamefont {Kamenev}, \citenamefont {Palevski}, \citenamefont {Pfeiffer},\
  and\ \citenamefont {West}}]{Berk1995}%
  \BibitemOpen
  \bibfield  {author} {\bibinfo {author} {\bibfnamefont {Y.}~\bibnamefont
  {Berk}}, \bibinfo {author} {\bibfnamefont {A.}~\bibnamefont {Kamenev}},
  \bibinfo {author} {\bibfnamefont {A.}~\bibnamefont {Palevski}}, \bibinfo
  {author} {\bibfnamefont {L.~N.}\ \bibnamefont {Pfeiffer}}, \ and\ \bibinfo
  {author} {\bibfnamefont {K.~W.}\ \bibnamefont {West}},\ }\href {\doibase
  10.1103/PhysRevB.51.2604} {\bibfield  {journal} {\bibinfo  {journal} {Phys.
  Rev. B}\ }\textbf {\bibinfo {volume} {51}},\ \bibinfo {pages} {2604}
  (\bibinfo {year} {1995})}\BibitemShut {NoStop}%
\bibitem [{\citenamefont {Murphy}\ \emph {et~al.}(1995)\citenamefont {Murphy},
  \citenamefont {Eisenstein}, \citenamefont {Pfeiffer},\ and\ \citenamefont
  {West}}]{Murphy1995}%
  \BibitemOpen
  \bibfield  {author} {\bibinfo {author} {\bibfnamefont {S.~Q.}\ \bibnamefont
  {Murphy}}, \bibinfo {author} {\bibfnamefont {J.~P.}\ \bibnamefont
  {Eisenstein}}, \bibinfo {author} {\bibfnamefont {L.~N.}\ \bibnamefont
  {Pfeiffer}}, \ and\ \bibinfo {author} {\bibfnamefont {K.~W.}\ \bibnamefont
  {West}},\ }\href {\doibase 10.1103/PhysRevB.52.14825} {\bibfield  {journal}
  {\bibinfo  {journal} {Phys. Rev. B}\ }\textbf {\bibinfo {volume} {52}},\
  \bibinfo {pages} {14825} (\bibinfo {year} {1995})}\BibitemShut {NoStop}%
\bibitem [{\citenamefont {Slutzky}\ \emph {et~al.}(1996)\citenamefont
  {Slutzky}, \citenamefont {Entin-Wohlman}, \citenamefont {Berk}, \citenamefont
  {Palevski},\ and\ \citenamefont {Shtrikman}}]{Slutzky1996}%
  \BibitemOpen
  \bibfield  {author} {\bibinfo {author} {\bibfnamefont {M.}~\bibnamefont
  {Slutzky}}, \bibinfo {author} {\bibfnamefont {O.}~\bibnamefont
  {Entin-Wohlman}}, \bibinfo {author} {\bibfnamefont {Y.}~\bibnamefont {Berk}},
  \bibinfo {author} {\bibfnamefont {A.}~\bibnamefont {Palevski}}, \ and\
  \bibinfo {author} {\bibfnamefont {H.}~\bibnamefont {Shtrikman}},\ }\href
  {\doibase 10.1103/PhysRevB.53.4065} {\bibfield  {journal} {\bibinfo
  {journal} {Phys. Rev. B}\ }\textbf {\bibinfo {volume} {53}},\ \bibinfo
  {pages} {4065} (\bibinfo {year} {1996})}\BibitemShut {NoStop}%
\bibitem [{\citenamefont {Bostwick}\ \emph {et~al.}(2008)\citenamefont
  {Bostwick}, \citenamefont {Speck}, \citenamefont {Seyller}, \citenamefont
  {Horn}, \citenamefont {Polini}, \citenamefont {Asgari}, \citenamefont
  {MacDonald},\ and\ \citenamefont {Rotenberg}}]{bostwick2010}%
  \BibitemOpen
  \bibfield  {author} {\bibinfo {author} {\bibfnamefont {A.}~\bibnamefont
  {Bostwick}}, \bibinfo {author} {\bibfnamefont {F.}~\bibnamefont {Speck}},
  \bibinfo {author} {\bibfnamefont {T.}~\bibnamefont {Seyller}}, \bibinfo
  {author} {\bibfnamefont {K.}~\bibnamefont {Horn}}, \bibinfo {author}
  {\bibfnamefont {M.}~\bibnamefont {Polini}}, \bibinfo {author} {\bibfnamefont
  {R.}~\bibnamefont {Asgari}}, \bibinfo {author} {\bibfnamefont {A.~H.}\
  \bibnamefont {MacDonald}}, \ and\ \bibinfo {author} {\bibfnamefont
  {E.}~\bibnamefont {Rotenberg}},\ }\href {\doibase 10.1126/science.1186489}
  {\bibfield  {journal} {\bibinfo  {journal} {Science}\ }\textbf {\bibinfo
  {volume} {328}},\ \bibinfo {pages} {999} (\bibinfo {year}
  {2008})}\BibitemShut {NoStop}%
\bibitem [{\citenamefont {Lee}(2018)}]{Lee2018}%
  \BibitemOpen
  \bibfield  {author} {\bibinfo {author} {\bibfnamefont {S.-S.}\ \bibnamefont
  {Lee}},\ }\href {\doibase
  https://doi.org/10.1146/annurev-conmatphys-031016-025531} {\bibfield
  {journal} {\bibinfo  {journal} {Annual Review of Condensed Matter Physics}\
  }\textbf {\bibinfo {volume} {9}} (\bibinfo {year} {2018}),\
  https://doi.org/10.1146/annurev-conmatphys-031016-025531}\BibitemShut
  {NoStop}%
\bibitem [{\citenamefont {Menashe}\ and\ \citenamefont
  {Laikhtman}(1996)}]{Menashe1996}%
  \BibitemOpen
  \bibfield  {author} {\bibinfo {author} {\bibfnamefont {D.}~\bibnamefont
  {Menashe}}\ and\ \bibinfo {author} {\bibfnamefont {B.}~\bibnamefont
  {Laikhtman}},\ }\href {\doibase 10.1103/PhysRevB.54.11561} {\bibfield
  {journal} {\bibinfo  {journal} {Phys. Rev. B}\ }\textbf {\bibinfo {volume}
  {54}},\ \bibinfo {pages} {11561} (\bibinfo {year} {1996})}\BibitemShut
  {NoStop}%
\bibitem [{\citenamefont {Seydi}\ \emph {et~al.}(2018)\citenamefont {Seydi},
  \citenamefont {Abedinpour}, \citenamefont {Asgari},\ and\ \citenamefont
  {Tanatar}}]{Seydi_PRA2018}%
  \BibitemOpen
  \bibfield  {author} {\bibinfo {author} {\bibfnamefont {I.}~\bibnamefont
  {Seydi}}, \bibinfo {author} {\bibfnamefont {S.~H.}\ \bibnamefont
  {Abedinpour}}, \bibinfo {author} {\bibfnamefont {R.}~\bibnamefont {Asgari}},
  \ and\ \bibinfo {author} {\bibfnamefont {B.}~\bibnamefont {Tanatar}},\ }\href
  {\doibase 10.1103/PhysRevA.98.063623} {\bibfield  {journal} {\bibinfo
  {journal} {Phys. Rev. A}\ }\textbf {\bibinfo {volume} {98}},\ \bibinfo
  {pages} {063623} (\bibinfo {year} {2018})}\BibitemShut {NoStop}%
\bibitem [{\citenamefont {Gerlach}\ \emph {et~al.}(2017)\citenamefont
  {Gerlach}, \citenamefont {Schattner}, \citenamefont {Berg},\ and\
  \citenamefont {Trebst}}]{Gerlach2017}%
  \BibitemOpen
  \bibfield  {author} {\bibinfo {author} {\bibfnamefont {M.~H.}\ \bibnamefont
  {Gerlach}}, \bibinfo {author} {\bibfnamefont {Y.}~\bibnamefont {Schattner}},
  \bibinfo {author} {\bibfnamefont {E.}~\bibnamefont {Berg}}, \ and\ \bibinfo
  {author} {\bibfnamefont {S.}~\bibnamefont {Trebst}},\ }\href {\doibase
  10.1103/PhysRevB.95.035124} {\bibfield  {journal} {\bibinfo  {journal} {Phys.
  Rev. B}\ }\textbf {\bibinfo {volume} {95}},\ \bibinfo {pages} {035124}
  (\bibinfo {year} {2017})}\BibitemShut {NoStop}%
\bibitem [{\citenamefont {Recati}\ and\ \citenamefont
  {Stringari}(2023)}]{Recati_NRP2023}%
  \BibitemOpen
  \bibfield  {author} {\bibinfo {author} {\bibfnamefont {A.}~\bibnamefont
  {Recati}}\ and\ \bibinfo {author} {\bibfnamefont {S.}~\bibnamefont
  {Stringari}},\ }\href {\doibase 10.1038/s42254-023-00648-2} {\bibfield
  {journal} {\bibinfo  {journal} {Nature Reviews Physics}\ }\textbf {\bibinfo
  {volume} {5}},\ \bibinfo {pages} {735–743} (\bibinfo {year}
  {2023})}\BibitemShut {NoStop}%
\bibitem [{\citenamefont {Han}\ \emph {et~al.}(2018)\citenamefont {Han},
  \citenamefont {Zhang}, \citenamefont {Wang}, \citenamefont {Jiang},
  \citenamefont {Zhang},\ and\ \citenamefont {Zhang}}]{Han_PRL2018}%
  \BibitemOpen
  \bibfield  {author} {\bibinfo {author} {\bibfnamefont {W.}~\bibnamefont
  {Han}}, \bibinfo {author} {\bibfnamefont {X.-F.}\ \bibnamefont {Zhang}},
  \bibinfo {author} {\bibfnamefont {D.-S.}\ \bibnamefont {Wang}}, \bibinfo
  {author} {\bibfnamefont {H.-F.}\ \bibnamefont {Jiang}}, \bibinfo {author}
  {\bibfnamefont {W.}~\bibnamefont {Zhang}}, \ and\ \bibinfo {author}
  {\bibfnamefont {S.-G.}\ \bibnamefont {Zhang}},\ }\href {\doibase
  10.1103/PhysRevLett.121.030404} {\bibfield  {journal} {\bibinfo  {journal}
  {Phys. Rev. Lett.}\ }\textbf {\bibinfo {volume} {121}},\ \bibinfo {pages}
  {030404} (\bibinfo {year} {2018})}\BibitemShut {NoStop}%
\bibitem [{\citenamefont {Zhu}\ \emph {et~al.}(2024)\citenamefont {Zhu},
  \citenamefont {Ma}, \citenamefont {Bai}, \citenamefont {Yu}, \citenamefont
  {Ye}, \citenamefont {Li}, \citenamefont {Zhuang},\ and\ \citenamefont
  {Liu}}]{Zhu_PRR2023}%
  \BibitemOpen
  \bibfield  {author} {\bibinfo {author} {\bibfnamefont {H.}~\bibnamefont
  {Zhu}}, \bibinfo {author} {\bibfnamefont {Y.-Q.}\ \bibnamefont {Ma}},
  \bibinfo {author} {\bibfnamefont {W.-K.}\ \bibnamefont {Bai}}, \bibinfo
  {author} {\bibfnamefont {Y.-M.}\ \bibnamefont {Yu}}, \bibinfo {author}
  {\bibfnamefont {F.-F.}\ \bibnamefont {Ye}}, \bibinfo {author} {\bibfnamefont
  {Y.-Y.}\ \bibnamefont {Li}}, \bibinfo {author} {\bibfnamefont
  {L.}~\bibnamefont {Zhuang}}, \ and\ \bibinfo {author} {\bibfnamefont {W.-M.}\
  \bibnamefont {Liu}},\ }\href {\doibase 10.1103/PhysRevResearch.6.023151}
  {\bibfield  {journal} {\bibinfo  {journal} {Phys. Rev. Res.}\ }\textbf
  {\bibinfo {volume} {6}},\ \bibinfo {pages} {023151} (\bibinfo {year}
  {2024})}\BibitemShut {NoStop}%
\bibitem [{\citenamefont {Song}\ \emph {et~al.}(2023)\citenamefont {Song},
  \citenamefont {Ma}, \citenamefont {Kallin},\ and\ \citenamefont
  {Lee}}]{Song2023}%
  \BibitemOpen
  \bibfield  {author} {\bibinfo {author} {\bibfnamefont {H.}~\bibnamefont
  {Song}}, \bibinfo {author} {\bibfnamefont {H.}~\bibnamefont {Ma}}, \bibinfo
  {author} {\bibfnamefont {C.}~\bibnamefont {Kallin}}, \ and\ \bibinfo {author}
  {\bibfnamefont {S.-S.}\ \bibnamefont {Lee}},\ }\href@noop {} {\enquote
  {\bibinfo {title} {Anomalous quasiparticle lifetime in geometric quantum
  critical metals},}\ } (\bibinfo {year} {2023}),\ \Eprint
  {http://arxiv.org/abs/2310.07539} {arXiv:2310.07539 [cond-mat.str-el]}
  \BibitemShut {NoStop}%
\bibitem [{\citenamefont {Holder}\ and\ \citenamefont
  {Metzner}(2014)}]{Holder2014}%
  \BibitemOpen
  \bibfield  {author} {\bibinfo {author} {\bibfnamefont {T.}~\bibnamefont
  {Holder}}\ and\ \bibinfo {author} {\bibfnamefont {W.}~\bibnamefont
  {Metzner}},\ }\href {\doibase 10.1103/PhysRevB.90.161106} {\bibfield
  {journal} {\bibinfo  {journal} {Phys. Rev. B}\ }\textbf {\bibinfo {volume}
  {90}},\ \bibinfo {pages} {161106(R)} (\bibinfo {year} {2014})}\BibitemShut
  {NoStop}%
\bibitem [{\citenamefont {L\"ohneysen}\ \emph {et~al.}(2007)\citenamefont
  {L\"ohneysen}, \citenamefont {Rosch}, \citenamefont {Vojta},\ and\
  \citenamefont {W\"olfle}}]{Lohneysen2007}%
  \BibitemOpen
  \bibfield  {author} {\bibinfo {author} {\bibfnamefont {H.~v.}\ \bibnamefont
  {L\"ohneysen}}, \bibinfo {author} {\bibfnamefont {A.}~\bibnamefont {Rosch}},
  \bibinfo {author} {\bibfnamefont {M.}~\bibnamefont {Vojta}}, \ and\ \bibinfo
  {author} {\bibfnamefont {P.}~\bibnamefont {W\"olfle}},\ }\href {\doibase
  10.1103/RevModPhys.79.1015} {\bibfield  {journal} {\bibinfo  {journal} {Rev.
  Mod. Phys.}\ }\textbf {\bibinfo {volume} {79}},\ \bibinfo {pages} {1015}
  (\bibinfo {year} {2007})}\BibitemShut {NoStop}%
\bibitem [{\citenamefont {Gr\"uner}(1988)}]{Gruner_RMP1988}%
  \BibitemOpen
  \bibfield  {author} {\bibinfo {author} {\bibfnamefont {G.}~\bibnamefont
  {Gr\"uner}},\ }\href {\doibase 10.1103/RevModPhys.60.1129} {\bibfield
  {journal} {\bibinfo  {journal} {Rev. Mod. Phys.}\ }\textbf {\bibinfo {volume}
  {60}},\ \bibinfo {pages} {1129} (\bibinfo {year} {1988})}\BibitemShut
  {NoStop}%
\bibitem [{\citenamefont {Zhu}\ and\ \citenamefont {Han}(2022)}]{Zhu_JPCS2022}%
  \BibitemOpen
  \bibfield  {author} {\bibinfo {author} {\bibfnamefont {H.}~\bibnamefont
  {Zhu}}\ and\ \bibinfo {author} {\bibfnamefont {H.}~\bibnamefont {Han}},\
  }\href {\doibase 10.1088/1742-6596/2338/1/012028} {\bibfield  {journal}
  {\bibinfo  {journal} {Journal of Physics: Conference Series}\ }\textbf
  {\bibinfo {volume} {2338}},\ \bibinfo {pages} {012028} (\bibinfo {year}
  {2022})}\BibitemShut {NoStop}%
\bibitem [{\citenamefont {B\"ohm}\ \emph {et~al.}(2010)\citenamefont {B\"ohm},
  \citenamefont {Holler}, \citenamefont {Krotscheck},\ and\ \citenamefont
  {Panholzer}}]{Bohm_PRB2010}%
  \BibitemOpen
  \bibfield  {author} {\bibinfo {author} {\bibfnamefont {H.~M.}\ \bibnamefont
  {B\"ohm}}, \bibinfo {author} {\bibfnamefont {R.}~\bibnamefont {Holler}},
  \bibinfo {author} {\bibfnamefont {E.}~\bibnamefont {Krotscheck}}, \ and\
  \bibinfo {author} {\bibfnamefont {M.}~\bibnamefont {Panholzer}},\ }\href
  {\doibase 10.1103/PhysRevB.82.224505} {\bibfield  {journal} {\bibinfo
  {journal} {Phys. Rev. B}\ }\textbf {\bibinfo {volume} {82}},\ \bibinfo
  {pages} {224505} (\bibinfo {year} {2010})}\BibitemShut {NoStop}%
\bibitem [{\citenamefont {Godfrin}\ \emph {et~al.}(2012)\citenamefont
  {Godfrin}, \citenamefont {Meschke}, \citenamefont {Lauter}, \citenamefont
  {Sultan}, \citenamefont {Böhm}, \citenamefont {Krotscheck},\ and\
  \citenamefont {Panholzer}}]{Godfrin2012}%
  \BibitemOpen
  \bibfield  {author} {\bibinfo {author} {\bibfnamefont {H.}~\bibnamefont
  {Godfrin}}, \bibinfo {author} {\bibfnamefont {M.}~\bibnamefont {Meschke}},
  \bibinfo {author} {\bibfnamefont {H.~J.}\ \bibnamefont {Lauter}}, \bibinfo
  {author} {\bibfnamefont {A.}~\bibnamefont {Sultan}}, \bibinfo {author}
  {\bibfnamefont {H.~M.}\ \bibnamefont {Böhm}}, \bibinfo {author}
  {\bibfnamefont {E.}~\bibnamefont {Krotscheck}}, \ and\ \bibinfo {author}
  {\bibfnamefont {M.}~\bibnamefont {Panholzer}},\ }\href {\doibase
  10.1038/nature10919} {\bibfield  {journal} {\bibinfo  {journal} {Nature}\
  }\textbf {\bibinfo {volume} {483}},\ \bibinfo {pages} {576} (\bibinfo {year}
  {2012})}\BibitemShut {NoStop}%
\bibitem [{\citenamefont {Khasseh}\ \emph {et~al.}(2017)\citenamefont
  {Khasseh}, \citenamefont {Abedinpour},\ and\ \citenamefont
  {Tanatar}}]{khasseh2017phase}%
  \BibitemOpen
  \bibfield  {author} {\bibinfo {author} {\bibfnamefont {R.}~\bibnamefont
  {Khasseh}}, \bibinfo {author} {\bibfnamefont {S.~H.}\ \bibnamefont
  {Abedinpour}}, \ and\ \bibinfo {author} {\bibfnamefont {B.}~\bibnamefont
  {Tanatar}},\ }\href {\doibase 10.1103/PhysRevA.96.053611} {\bibfield
  {journal} {\bibinfo  {journal} {Phys. Rev. A}\ }\textbf {\bibinfo {volume}
  {96}},\ \bibinfo {pages} {053611} (\bibinfo {year} {2017})}\BibitemShut
  {NoStop}%
\bibitem [{\citenamefont {Seydi}\ \emph {et~al.}(2021)\citenamefont {Seydi},
  \citenamefont {Abedinpour}, \citenamefont {Asgari}, \citenamefont
  {Panholzer},\ and\ \citenamefont {Tanatar}}]{Seydi2021}%
  \BibitemOpen
  \bibfield  {author} {\bibinfo {author} {\bibfnamefont {I.}~\bibnamefont
  {Seydi}}, \bibinfo {author} {\bibfnamefont {S.~H.}\ \bibnamefont
  {Abedinpour}}, \bibinfo {author} {\bibfnamefont {R.}~\bibnamefont {Asgari}},
  \bibinfo {author} {\bibfnamefont {M.}~\bibnamefont {Panholzer}}, \ and\
  \bibinfo {author} {\bibfnamefont {B.}~\bibnamefont {Tanatar}},\ }\href
  {\doibase 10.1103/PhysRevA.103.043308} {\bibfield  {journal} {\bibinfo
  {journal} {Phys. Rev. A}\ }\textbf {\bibinfo {volume} {103}},\ \bibinfo
  {pages} {043308} (\bibinfo {year} {2021})}\BibitemShut {NoStop}%
\bibitem [{Not()}]{Note2}%
  \BibitemOpen
  \href@noop {} {}\bibinfo {note} {See Supplemental Material at (URL will be
  placed by publisher) for the details.}\BibitemShut {Stop}%
\bibitem [{\citenamefont {Khodel}\ \emph {et~al.}(1997)\citenamefont {Khodel},
  \citenamefont {Shaginyan},\ and\ \citenamefont
  {Zverev}}]{Khodel_JETPLett1997}%
  \BibitemOpen
  \bibfield  {author} {\bibinfo {author} {\bibfnamefont {V.~A.}\ \bibnamefont
  {Khodel}}, \bibinfo {author} {\bibfnamefont {V.~R.}\ \bibnamefont
  {Shaginyan}}, \ and\ \bibinfo {author} {\bibfnamefont {M.~V.}\ \bibnamefont
  {Zverev}},\ }\href {\doibase 10.1134/1.567356} {\bibfield  {journal}
  {\bibinfo  {journal} {Journal of Experimental and Theoretical Physics
  Letters}\ }\textbf {\bibinfo {volume} {65}},\ \bibinfo {pages} {253–258}
  (\bibinfo {year} {1997})}\BibitemShut {NoStop}%
\bibitem [{\citenamefont {Yakovenko}\ and\ \citenamefont
  {Khodel}(2003)}]{Yakovenko_JETPLett2003}%
  \BibitemOpen
  \bibfield  {author} {\bibinfo {author} {\bibfnamefont {V.~M.}\ \bibnamefont
  {Yakovenko}}\ and\ \bibinfo {author} {\bibfnamefont {V.~A.}\ \bibnamefont
  {Khodel}},\ }\href {\doibase 10.1134/1.1630135} {\bibfield  {journal}
  {\bibinfo  {journal} {Journal of Experimental and Theoretical Physics
  Letters}\ }\textbf {\bibinfo {volume} {78}},\ \bibinfo {pages} {398–401}
  (\bibinfo {year} {2003})}\BibitemShut {NoStop}%
\bibitem [{\citenamefont {Zhang}\ \emph {et~al.}(2005)\citenamefont {Zhang},
  \citenamefont {Yakovenko},\ and\ \citenamefont {Das~Sarma}}]{Zhang_PRB2005}%
  \BibitemOpen
  \bibfield  {author} {\bibinfo {author} {\bibfnamefont {Y.}~\bibnamefont
  {Zhang}}, \bibinfo {author} {\bibfnamefont {V.~M.}\ \bibnamefont
  {Yakovenko}}, \ and\ \bibinfo {author} {\bibfnamefont {S.}~\bibnamefont
  {Das~Sarma}},\ }\href {\doibase 10.1103/PhysRevB.71.115105} {\bibfield
  {journal} {\bibinfo  {journal} {Phys. Rev. B}\ }\textbf {\bibinfo {volume}
  {71}},\ \bibinfo {pages} {115105} (\bibinfo {year} {2005})}\BibitemShut
  {NoStop}%
\bibitem [{\citenamefont {Asgari}\ \emph {et~al.}(2009)\citenamefont {Asgari},
  \citenamefont {Gokmen}, \citenamefont {Tanatar}, \citenamefont
  {Padmanabhan},\ and\ \citenamefont {Shayegan}}]{Asgari_PRB2009}%
  \BibitemOpen
  \bibfield  {author} {\bibinfo {author} {\bibfnamefont {R.}~\bibnamefont
  {Asgari}}, \bibinfo {author} {\bibfnamefont {T.}~\bibnamefont {Gokmen}},
  \bibinfo {author} {\bibfnamefont {B.}~\bibnamefont {Tanatar}}, \bibinfo
  {author} {\bibfnamefont {M.}~\bibnamefont {Padmanabhan}}, \ and\ \bibinfo
  {author} {\bibfnamefont {M.}~\bibnamefont {Shayegan}},\ }\href {\doibase
  10.1103/PhysRevB.79.235324} {\bibfield  {journal} {\bibinfo  {journal} {Phys.
  Rev. B}\ }\textbf {\bibinfo {volume} {79}},\ \bibinfo {pages} {235324}
  (\bibinfo {year} {2009})}\BibitemShut {NoStop}%
\bibitem [{\citenamefont {Seydi}\ and\ \citenamefont
  {et~al.}(2024)}]{Iran_quasiparticle_draft}%
  \BibitemOpen
  \bibfield  {author} {\bibinfo {author} {\bibfnamefont {I.}~\bibnamefont
  {Seydi}}\ and\ \bibinfo {author} {\bibnamefont {et~al.}},\ }\href@noop {}
  {\enquote {\bibinfo {title} {In preparation},}\ } (\bibinfo {year}
  {2024})\BibitemShut {NoStop}%
\end{thebibliography}%

\newpage
\pagebreak
\clearpage
\widetext

\begin{center}
	\textbf{\large Supplemental Material for ``Anomalous Lifetime of Quasiparticles in Fermi Liquids as a Precursor of the Density-Wave Instability"}
\end{center}

\setcounter{equation}{0}
\setcounter{figure}{0}
\setcounter{table}{0}
\setcounter{page}{1}
\makeatletter
\renewcommand{\theequation}{S\arabic{equation}}
\renewcommand{\thefigure}{S\arabic{figure}}

\section{The imaginary part of self-energy within the on-shell approximation}\label{sec:Im_E}
The imaginary part of G$_0$W self-energy is given by
\be\label{sigma_im}
{\rm Im}\,\Sigma (k,E) = \int \frac{\mathrm{d}{\bf q}}{(2\pi)^3} v^{2}(q){\rm Im}\,\chi(q,E-\xi_{|\textbf{k}-\textbf{q}|}) 
\left[\Theta(E-\xi_{|\textbf{k}-\textbf{q}|})-\Theta(-\xi_{|\textbf{k}-\textbf{q}|})\right].
\ee
Here, $\xi_k=\varepsilon_k-\varepsilon_{\rm F}$, and $\varepsilon_{\rm F}=\hbar^2 k^2_{\rm F}/(2 m)$ is the non-interacting Fermi energy. 
Within the on-shell approximation i.e., $E\to \xi_k$, and making a change of variables $\pv=\kv-\qv$, we find
\be\label{sigma_im_OSA}
\begin{split}
	{\rm Im}\,\Sigma (k,\xi_k) &=  \int \frac{\mathrm{d}^3{\bf p}}{(2\pi)^3} v^{2}(|\kv-\pv|){\rm Im }\,\chi(|\kv-\pv|,\xi_k-\xi_{p}) 
	\left[\Theta(\xi_k-\xi_p)-\Theta(-\xi_p)\right]\\
	&= {\rm sgn}(k-k_{\rm F}) \int_{k_{\rm F}}^k \mathrm{d}p \,p^{2} \int \frac{\mathrm{d}\Omega}{(2\pi)^3} \, v^{2}(|\kv-\pv|){\rm Im }\,\chi(|\kv-\pv|,\xi_k-\xi_{p}),
\end{split}
\ee
where  $\mathrm{d}\Omega=\sin\theta\mathrm{d}\theta\mathrm{d}\phi$ is the 3D solid angle element.
From now on, we use the dimensionless quantities, where all the energies are written in the units of $\varepsilon_{\rm F}$, all the wave-vectors in the units of $k_{\rm F}$, response functions in the units of the density-of-states per unit volume at the Fermi surface (DOS), and interaction $v_q$ in the units of inverse DOS. 
Note that the DOS of a 3D system is $\nu_0=m k_{\rm F}/(2\pi^2\hbar^2)$.

Now, the  dimensionless self-energy reads
\be\label{sigma_im_OSA2}
\begin{split}
	{\rm Im }\,\Sigma (k,\xi_k)
	= & \int_{1}^k \mathrm{d}p \int_{-1}^{1} \mathrm{d}\cos\theta\, p^2v^{2}(|\kv-\pv|){\rm Im }\,\chi(|\kv-\pv|,k^2-p^2)\\
	=&\int_{1}^k \mathrm{d}p\frac{ p}{k} \int_{k-p}^{k+p} \mathrm{d}y \,y\, v^{2}(y)\frac{{\rm Im }\,\chi_0(y,k^2-p^2)}{\left|1-v(y) \chi_0(y,k^2-p^2)\right|^2}\\
	=& -\frac{\pi}{4} \int_{1}^k \mathrm{d}p \frac{ p(k^2-p^2)}{k} \int_{k-p}^{k+p} \mathrm{d}y \frac{ v^{2}(y)}{\left|1-v(y) \chi_0(y,k^2-p^2)\right|^2}.
\end{split}
\ee
In the second line, we have used $y=\sqrt{k^2+p^2- 2 kp \cos\theta}$, and in the third line we have made use of~\cite{Vignale2005}
\be\label{eq:im_chi0}
{\rm Im }\,\chi_0(0<q< 2,E\to 0)=- \frac{\pi}{4}\frac{E}{q},
\ee
in the dimensionless units. 

The leading order behavior of the self-energy close to the Fermi energy (i.e., $k\to 1$, limit) depends on the behavior of the denominator, i.e., the dielectric function $1-v(q)\chi_{0} (q, E)$ in the static $E\to 0$ limit. 
Below, we will briefly discuss the details of the calculations of three different regimes.
\section{The behavior of inverse lifetime close to the Fermi level}\label{app:self}
\subsection{Normal Fermi liquid phase}
For normal Fermi liquids, the static dielectric function $1-v(q)\chi_{0} (q)$ never becomes zero (in fact, it is always positive). Therefore, the $k\to1$ limit could be safely taken in the second integrand in Eq.~\eqref{sigma_im_OSA2}
\be\label{sigma_im_FL_app}
\begin{split}
	{\rm Im }\,\Sigma (k\to 1,\xi_k\to 0)
	=& -\frac{\pi}{4 k} \int_{1}^k \mathrm{d}p\, p(k^2-p^2) \int_{0}^{2} \mathrm{d}y \frac{ v^{2}(y)}{\left|1-v(y) \chi_0(y)\right|^2}\\
	\approx & -\left(\frac{\pi}{4} \int_{0}^{2} \mathrm{d}y \frac{ v^{2}(y)}{\left|1-v(y) \chi_0(y)\right|^2}\right) \delta^2,
\end{split}
\ee
where $\delta=k-1$.
\subsection{Density-wave instability point}
If the static dielectric function of a homogeneous Fermi liquid vanishes at the wave vector $q_c$, i.e.,
\be\label{eq:epsilon_qc}
\varepsilon(q_c,0)=1-v(q_c) \chi_{0} (q_c,0)=0,
\ee
the homogeneous system becomes unstable to a density-modulated phase with wavelength $\lambda_c=2\pi/q_c$. 
The instability wave vector is usually close to the Fermi wave vector. In the following, we just require $0<q_c<2 k_{\rm F}$ (or $0<q_c< 2$, in the dimensionless units), to ensure a finite ${\rm Im}\, \chi_0(q\approx q_c, E \approx 0)$. 

In this instability regime, the integrand in Eq.~\eqref{sigma_im_OSA2} has a singularity in the $k\to 1$ limit, and the integral in the last line of Eq.~\eqref{sigma_im_FL_app} diverges. 
If we expand the dielectric function to leading orders around $q \approx q_c$ and $E \approx 0$, we find
\be\label{epsilon_expansion}
\begin{split}
	\varepsilon(q,E)\approx&1-v(q_c) \chi_{0} (q_c,0)\\
	&-\partial_q\left.[v(q) \chi_{0} (q,0)]\right|_{q=q_c}(q-q_c)\\
	&-\frac{1}{2}\partial^2_q\left.[v(q) \chi_{0} (q,0)]\right|_{q=q_c}(q-q_c)^2\\
	&-v(q_c)\partial_E\left.{\rm Re}\, \chi_{0} (q_c,E)\right|_{E=0}E\\
	&-i v(q_c)\partial_E\left.{\rm Im}\, \chi_{0} (q_c,E)\right|_{E=0}E.
\end{split}
\ee
The first line on the right-hand-side (RHS) of Eq.~\eqref{epsilon_expansion} vanishes according to Eq.~\eqref{eq:epsilon_qc}, the second line is zero, as it is the first derivative of a function at its minimum point, the fourth line vanishes as we have ${\rm Re}\, \chi_0(q, E\to 0)= \chi_0(q, 0)+{\cal O}(\omega^2)$~\cite{Vignale2005}. 
Finally, if we make use of the fact that ${\rm Im}\, \chi_0(q,\omega \to 0)$ is a linear function of $E$ [see, Eq.~\eqref{eq:im_chi0}], we find
\be
\epsilon(q\approx  q_c,E \approx 0) \approx 
-\frac{1}{2}\partial^2_q\left.[v(q) \chi_{0} (q)]\right|_{q=q_c}(q-q_c)^2
-i v(q_c) {\rm Im}\, \chi_{0} (q_c,E).
\ee
Now, it is easy to notice $|\varepsilon(y,k^2-p^2)|^2$ has four roots in the complex wave vector plane 
\be
z_j=q_c\pm (1\pm i)\sqrt{\left|{b}/{a}\right|},
\ee
with $a\equiv \partial^2_q\left.[v(q) \chi_{0} (q)]\right|_{q_c}$, and $b\equiv v(q_c) {\rm Im }\, \chi_{0} (q_c,k^2-p^2)=-\pi v(q_c)(k^2-p^2)/(4 q_c)$. Notice that the two roots are in the upper half of the complex plane, and the other two are in the lower half-plane. In the $k\to 1$ (and therefore $p\to 1$) limit, all four roots collapse at the $q_c$ point on the real axis. 
Now, the contribution of the singular term to the self-energy reads
\be\label{sigma_im_DWI}
\begin{split}
	{\rm Im}\,\Sigma (k,\xi_k)
	&= -\frac{\pi}{k a^2} \int_{1}^k \mathrm{d}p\, p(k^2-p^2) 
	\int_{k-p}^{k+p} \mathrm{d}y \frac{ v^{2}(y)}{(y-z_1)(y-z_2)(y-z_3)(y-z_4)}\\
	&= -\frac{\pi^2 v^2(q_c) }{4 \sqrt{|a|}} \left(\frac{4 q_c}{\pi v(q_c)}\right)^{3/2}\int_{1}^k \mathrm{d}p\, \frac{p(k^2-p^2)}{k(k^2-p^2)^{3/2}}\\
	&\approx- \sqrt{\frac{8 \pi q_c^3 v(q_c)}{\partial^2_q\left.[v(q) \chi_{0} (q)]\right|_{q_c}}}\,\delta^{1/2}.
\end{split}
\ee
In the last line, we have used Cauchy's residue theorem to evaluate the integral.

\subsection{Density-wave phase}
As we pass the instability point to the density-wave phase, the static dielectric function of the homogeneous phase becomes negative for $q_{c_1}<q<q_{c_2}$ (see Fig. 1 in the main text). Close to the phase transition point, these crossing wave vectors are expected to be smaller than $2k_{\rm F}$. Therefore, the integrand in Eq.~\eqref{sigma_im_OSA2} has two singularities. The expansion of the dielectric function close to each of these singularities is similar to Eq.~\eqref{epsilon_expansion}. However, the second line (first-order derivative with respect to $q$) does not vanish at the singular points.
We approximate the dielectric function close to the singular points as
\be\label{epsilon_expansion_lin}
\begin{split}
	\varepsilon(q\to q_{c,j},\xi_k)\approx
	-\partial_q\left.[v(q) \chi_{0} (q)]\right|_{q=q_{c,j}}(q-q_{c,j})-i v(q_{c,j}){\rm Im}\, \chi_{0} (q_{c,j},\xi_k).
\end{split}
\ee
The two poles of $|\varepsilon(y\sim q_{c,j},k^2-p^2)|^2$ in the complex domain are
\be
z_{j,\pm}=q_{c,j} \pm i |b_j/c_j|,
\ee
where $b\equiv v(q_{c,j}){\rm Im}\, \chi_{0} (q_{c,j},k^2-p^2)$, and $c_j\equiv \left.\partial_q[v(q) \chi_{0} (q)]\right|_{q_{c,j}}$, and only one pole around each $q_j$ falls in the upper half-plane.
After performing the integral over $\kappa$ using the residue theorem, we find
\be\label{sigma_im_BDWI}
\begin{split}
	{\rm Im}\,\Sigma (k\to 1,\xi_k\to 0)
	&= -\frac{\pi}{4} \sum_{j=1,2} \int_{1}^k \mathrm{d}p \frac{p(k^2-p^2)}{k |c_j^2}
	\int_{k-p}^{k+p}\mathrm{d}y \frac{v^{2}(y)}{(y-z_{j,+})(y-z_{j,-})}\\
	&= -\frac{\pi^2}{4} \sum_{j=1,2} v^2(q_{c,j})\int_{1}^k \mathrm{d}p \frac{p(k^2-p^2)}{k |c_j b_j|}\\
	& \approx- \pi  \sum_{j=1,2} \left|\frac{q_{c,j}v(q_{c,j})}{\left.\partial_q[v(q) \chi_{0} (q)]\right|_{q_{c,j}}} \right|\delta.
\end{split}
\ee

\section{Absence of collective-mode contribution to the quasiparticle lifetime}\label{app:collective}
Outside the particle-hole continuum, the imaginary part of the interacting density-density response function has a non-zero contribution right at the collective mode dispersion, where $1-w(q){\rm Re}\,\chi_{0} (q,\omega)=0$, i.e.,~\cite{Vignale2005}
\be\label{eq:im_chi_collec}
\left.	{\rm Im}\, \chi (q,\omega)\right|_{\Omega_{\rm col.}}\approx -\frac{\pi}{2}\frac{\Omega_{\rm col.}(q)}{w(q)}\delta\left(\omega-\Omega_{\rm col.}(q)\right).
\ee
Upon the substitution of this expression in Eq.~\eqref{sigma_im} and taking the on-shell approximation i.e, $E= \xi_k$, we find
\be\label{sigma_im_col}
\begin{split}
	\left.{\rm Im}\,\Sigma (k,\xi_k)\right|_{\rm col.} = -\frac{\pi\hbar}{2}\int \frac{\mathrm{d}{\bf q}}{(2\pi)^3} &v (q)\Omega_{\rm col.}(q)
	\delta\left(\xi_k-\xi_{|\textbf{k}-\textbf{q}|}-\hbar\Omega_{\rm col.}(q)\right)\\
	&\times \left[\Theta(\xi_k-\xi_{|\textbf{k}-\textbf{q}|})-\Theta(-\xi_{|\textbf{k}-\textbf{q}|})\right].
\end{split}
\ee
At the neighborhood of the Fermi surface, we take $k \to k_{\rm F}$ limit, which corresponds to $\xi_k\to 0$ too, and we can expand the Heaviside step function on the right-hand side of the equation \eqref{sigma_im_col}, as $\Theta(x+\epsilon)\approx \Theta(x)+\epsilon \delta(x)$ and find
\be\label{sigma_im_col_exp}
\begin{split}
	\left.{\rm Im }\,\Sigma (k,\xi_k)\right|_{\rm col.} &=- \frac{\pi \hbar \xi_k}{2}\int \frac{\mathrm{d}{\bf q}}{(2\pi)^3} v (q)\Omega_{\rm col.}(q)  \delta( \xi_{|\textbf{k}-\textbf{q}|})
	\delta\left(\xi_k-\xi_{|\textbf{k}-\textbf{q}|}-\hbar\Omega_{\rm col.}(q)\right)\\
	&=- \frac{\pi \hbar \xi_k}{2}\int \frac{\mathrm{d}{\bf q}}{(2\pi)^3} v (q)\Omega_{\rm col.}(q)  \delta( \xi_{|\textbf{k}-\textbf{q}|})
	\delta\left(\xi_k-\hbar\Omega_{\rm col.}(q)\right),
\end{split}
\ee
where in the second line, the first delta function is used to drop $\xi_{|\textbf{k}-\textbf{q}|}$ from the second delta function.
The collective density mode with short-range interaction (i.e, $v(q\to0)=w_0$), is sound wave with $\Omega_{\rm col.}(q\to 0)\approx v_s q$, where $v_s$ is the sound velocity. 
As undamped sound waves require $v_s>v_{\rm F}$, we can easily verify that satisfying both Dirac delta functions is impossible. Therefore, quasiparticles close to the Fermi surface can not decay into zero-sounds.

\section{Quasiparticle energy}\label{sec:EQP}
Interactions between particles in Fermi liquids renormalize the dynamical properties of quasiparticles, such as energy dispersion and effective mass. The real part of self-energy characterizes these renormalizations.

\begin{figure}
	\centering
	\includegraphics[width=\linewidth]{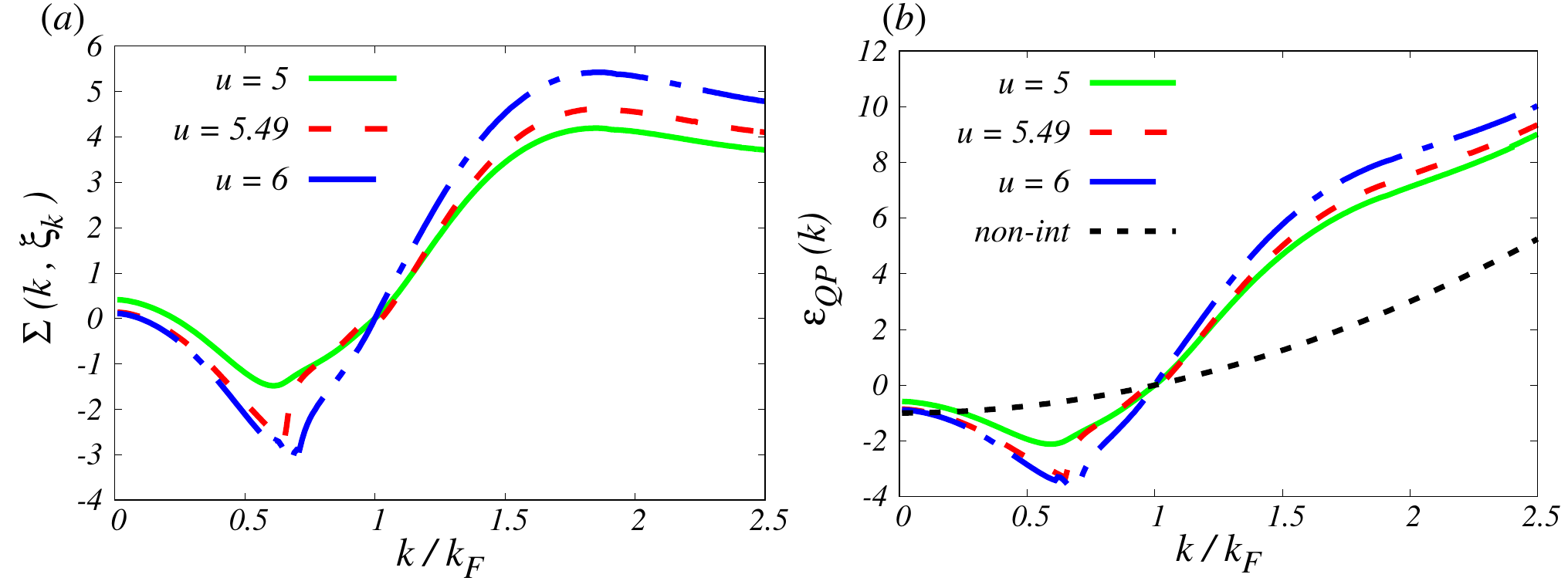} 	
	\caption{The real part of self-energy (left) and total quasiparticle energy (right), in the units of  $\epsilon_{\rm F}$, as a function of wave vector obtained in the on-shell approximation for 3D Rydberg-dressed Fermi liquid at different values of the dimensionless coupling strength $u$ for $r_c = 4$.
		The dashed line in the right panel shows the non-interacting energy.
		Note that at $r_c=4$, the system is in the normal Fermi liquid phase for $u=5$, the DWI appears at $u\approx 5.49$, and it is in the DW phase for $u=6$.
		\label{fig:E}} 
\end{figure}

Fig.~\ref{fig:E} shows the real part of self-energy and the total quasiparticle energy as a function of wave vector obtained in the on-shell approximation for 3D Rydberg-dressed Fermi liquid.  
At intermediate and high coupling strengths, the dynamic term of self-energy dominates the static (i.e., Hartree-Fock) part, and the quasiparticle energy is strongly different from the non-interacting dispersion.
 

\end{document}